\documentclass{JAC2003}

\usepackage{graphicx}

\begin{document}

\title{\vspace*{-1.8cm}
\begin{flushright}
{\bf\large LAL/RT 06-06}\\
{\normalsize\bf EUROTeV-Report-2006-038}\\
{\large June 2006}
\end{flushright}
\vspace*{1,5cm}
BENCHMARKING OF TRACKING CODES (BDSIM/DIMAD) 
USING THE ILC EXTRACTION LINES
\thanks{Work supported by the Commission of the European Communities
under the 6$^{th}$ Framework Programme "Structuring the European
\mbox{Research} Area", contract number RIDS-011899.}}

\author{{\bf\large R. Appleby,}\\ 
{\it The Cockcroft Institute and the University of Manchester,}\\ 
{\it Oxford Road, Manchester, M13 9PL, England
\vspace*{0,3cm}}\\
{\bf\large P. Bambade, O. Dadoun,}\\ 
{\it Laboratoire de l'Acc\'el\'erateur Lin\'eaire,}\\ 
{\it IN2P3-CNRS et Universit\'e de Paris-Sud 11, BP 34, 91898 Orsay cedex, France\vspace*{0,3cm}}\\
{\bf\large A. Ferrari\thanks{ferrari@tsl.uu.se}},\\
{\it Department of Nuclear and Particle Physics,}\\
{\it  Box 535, Uppsala University, 751 21 Uppsala, Sweden}
\vspace*{5cm}
}

\maketitle

\begin{abstract}
The study of beam transport is of central importance to the design and 
performance assessment of modern particle accelerators. In this work, we 
benchmark two contemporary codes - DIMAD and BDSIM, the latter being a 
\mbox{relatively} new tracking code built within the framework of GEANT4. 
We consider both the 20~mrad and 2~mrad extraction lines of the International 
Linear Collider (ILC) and we perform tracking studies of heavily disrupted 
post-collision electron beams. We find that the two codes mostly give an 
equivalent description of the beam transport.
\end{abstract}

\section{INTRODUCTION}

In a high-energy $e^+e^-$ linear collider such as ILC~\cite{ilc}, the 
beams must be focused to extremely small spot sizes in order to achieve 
high charge densities and, in turn, to reach the desired luminosity. This 
leads to large angular divergence and energy spread for the disrupted beams, 
as well as to the emission of beamstrahlung photons. A careful design 
of the extraction lines must therefore be performed to transport these 
outgoing beams from the interaction point to their dumps with small 
losses. At ILC, two configurations are being studied for the crossing 
angle at the interaction point and, as a result, for the design of the 
post-collision lines. With a 2~mrad crossing angle, the main challenge 
is the extraction of the disrupted beam, which has to be achieved by 
sending the outgoing beam off-center in the first large super-conducting 
defocusing quadrupole of the final focus beam line, as well as in the two 
nearby sextupoles~\cite{yuri}. On the other hand, with a 20~mrad crossing 
angle~\cite{ilc20a}, one must deal with technical difficulties such as large 
crab-crossing corrections or the construction of compact super-conducting 
quadrupoles for the incoming beam lines, as well as with the passage of 
the beams through the solenoid field with an angle. For the design of the 
ILC extraction lines, it is essential to have a reliable simulation program 
for particle tracking. In this study, we present a comparison between two 
codes, DIMAD~\cite{dimad} and BDSIM~\cite{bdsim}, using the present versions 
of the ILC post-collision lines for benchmarking purposes, in both large 
and small crossing angle cases.\\

The DIMAD program specifically aims at studying the behaviour of
particles in beam lines, by computing their trajectories using the 
second order matrix formalism~\cite{matrix}. The present version 
of the code ensures that the matrix treatment is correct to all 
orders for energy deviations~\cite{dimad}. This is important, as 
the ILC disrupted beams downstream of the interaction point can 
have very large energy spreads. As for the BDSIM program~\cite{bdsim},  
it uses the closed solutions in linear elements, whilst for higher-order 
elements, a GEANT4-like stepper integration method is used. The program 
is written in GEANT4~\cite{geant4} and provides a toolkit to simulate
interactions of particles with the surrounding matter once they have left
the vacuum pipe. However, for the purpose of this study, we only aim at 
comparing the tracking procedures in DIMAD and BDSIM.

\section{DIMAD AND BDSIM TRACKING IN THE 20~MRAD EXTRACTION LINE}

\begin{figure}[h!]
\centering
\includegraphics*[angle=-90,width=85mm]{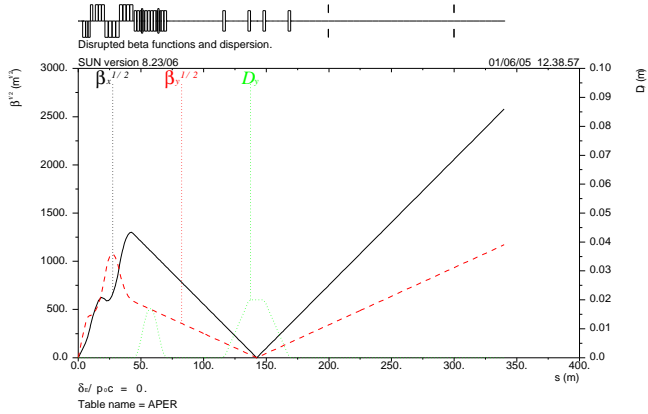}
\vspace*{0,2cm}
\caption{\it Betatron functions and vertical dispersion along the ILC extraction 
line with a 20~mrad crossing angle.}
\label{ilc20optics}
\end{figure}

For this benchmarking study, we use the nominal beam parameters 
for ILC at 500~GeV, for which the beam transport from 
the interaction point to the dump is almost loss-free (at least with 
the low-statistics input files that we use). The particle distributions 
for the disrupted beams at the interaction point were obtained 
with GUINEA-PIG~\cite{guinea} and then transported, with DIMAD or BDSIM, 
along the ILC 20~mrad extraction line. The optics used for this study 
consists of a DFDF quadruplet, followed by two vertical chicanes for 
energy and polarization measurements and a long field-free region that 
allows the beam to grow naturally, with two round collimators to reduce 
the maximum beam size at the dump, see Figure~\ref{ilc20optics}.\\ 

\begin{figure}[h!]
\centering
\includegraphics*[width=85mm, height=80mm]{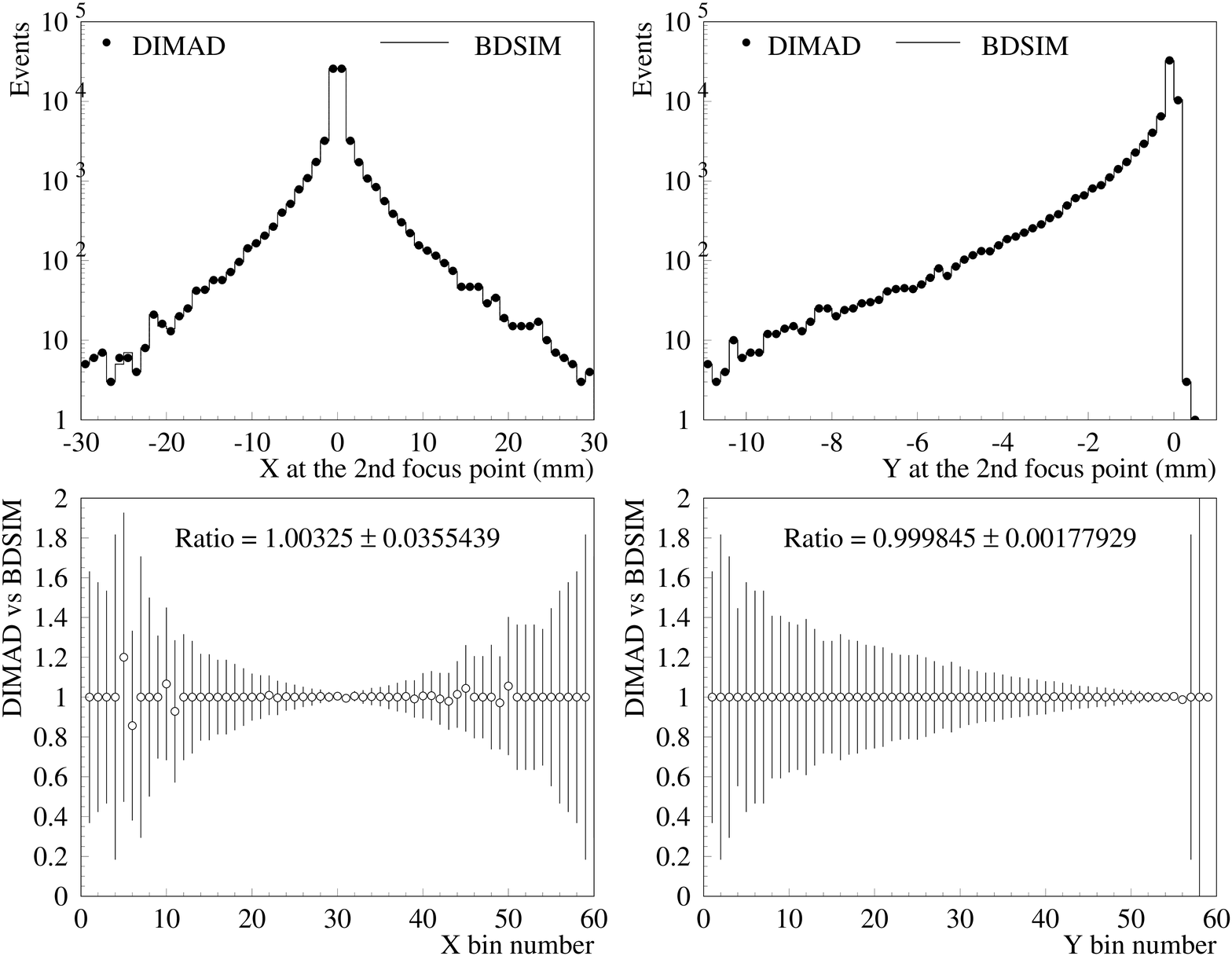}
\caption{\it Comparison of the transverse beam distributions obtained with DIMAD 
(full circles) and BDSIM (full line) at the secondary focus point 
Mexfoc. Both upper plots are distributed over 60 bins. The bottom plots 
show the ratio of the DIMAD and BDSIM distributions (see text for details).}
\label{focilc20}
\end{figure}

\begin{figure}[h!]
\centering
\includegraphics*[width=85mm, height=80mm]{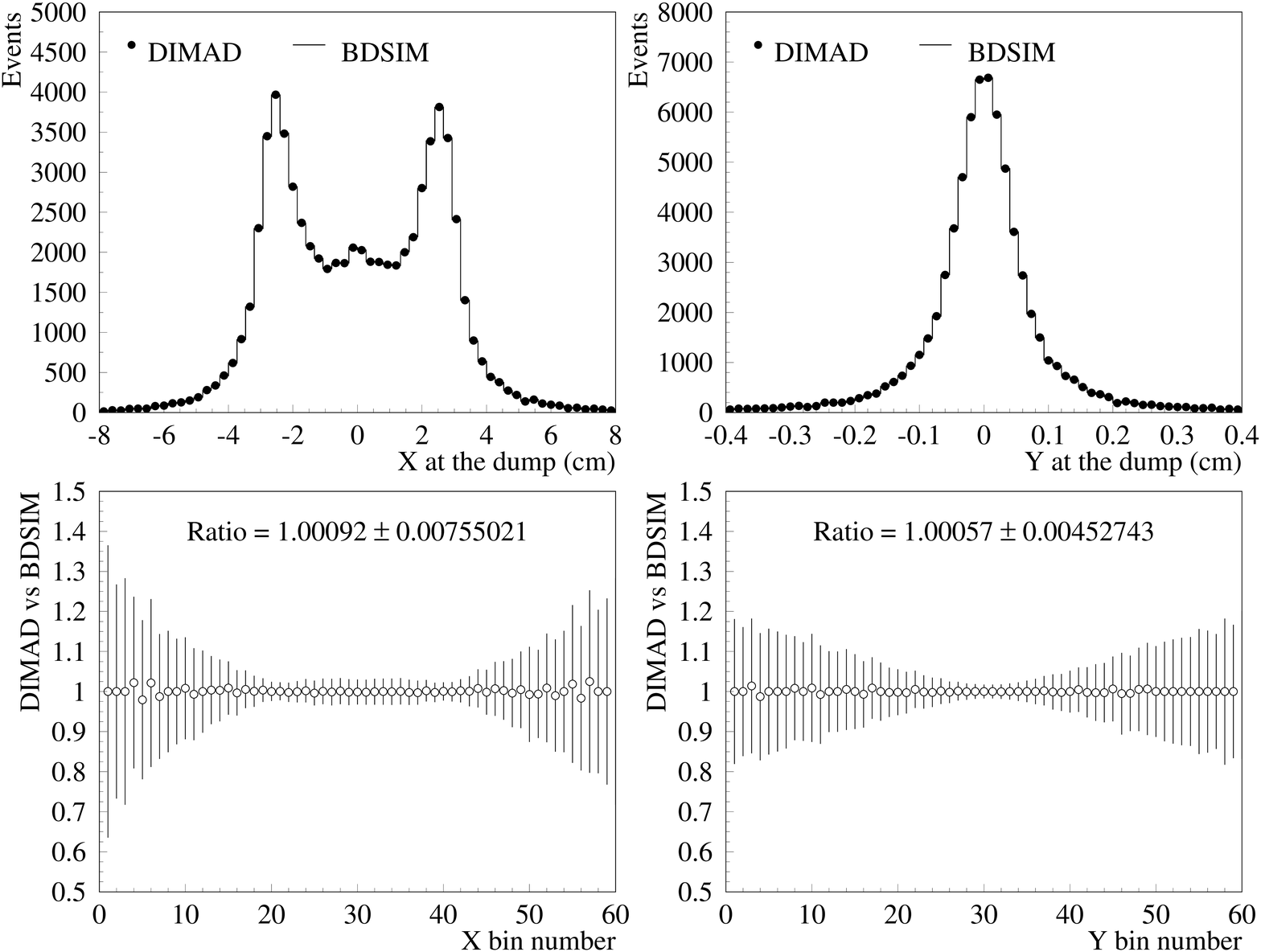}
\caption{\it Same as Figure~\ref{focilc20}, obtained at the end of the 20~mrad 
extraction line.}
\label{dumpilc20}
\end{figure}

At several locations of interest, we project the transverse beam 
distributions obtained with each program into binned histograms 
and we compare them quantitatively. An illustration of this procedure 
is shown in Figures~\ref{focilc20} and~\ref{dumpilc20} for the transverse 
beam distributions obtained respectively at the secondary focus point 
Mexfoc, located at $s = 142.4~\mbox{m}$ (where $\beta_x$ and $\beta_y$ 
are very small, with a vertical dispersion of 2~cm) and at the dump. The 
open circles show the ratio between the number of events found by DIMAD 
or BDSIM in a given histogram bin and the error bars account for the 
limited number of events per bin (very few events are found in the tails, 
which explains the large error bars there). The transverse distributions 
of the disrupted beams were also computed with DIMAD and BDSIM at several 
other locations in the 20~mrad extraction line. Their comparison also 
showed excellent agreement.

\vspace*{0,5cm}
\section{DIMAD AND BDSIM TRACKING IN THE 2~MRAD EXTRACTION LINE}

When the colliding beams cross with a small horizontal angle of 2~mrad, 
the outgoing disrupted beam passes off-axis through the bore of the final 
quadrupole QD0, both final sextupoles, but not the second-to-final quadrupole 
QF1. However, the outgoing beam sees the pocket field of this latter magnet. 
Following this doublet, the beam is focused with a series of large bore 
quadrupoles. These are followed by a vertical energy clean-up chicane, 
diagnostic chicanes for the purpose of energy spectrometry and 
polarimetry, and finally a long field-free region to allow the beam 
to grow to the dump, in the same way as in the 20~mrad scheme. Note 
that, in the 2~mrad scheme, separate dumps are used for the charged 
beam and for the beamstrahlung photons.
\vspace*{-0,2cm}

\begin{figure}[h!]
\centering
\hspace*{-0,9cm}\includegraphics*[angle=-90,width=98mm]{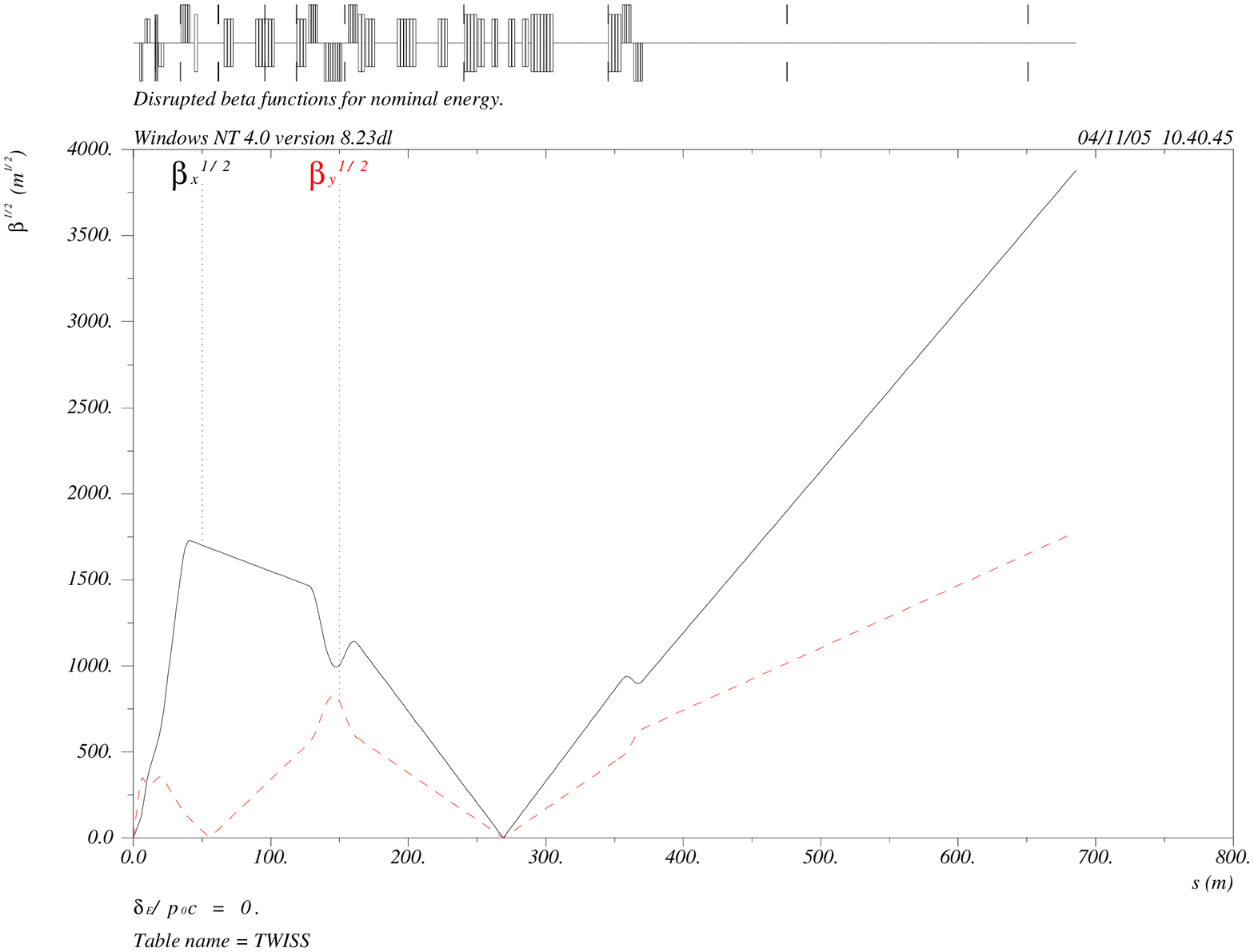}
\vspace*{-0,5cm}
\hspace*{-2,5cm}\includegraphics*[angle=-90,width=110mm]{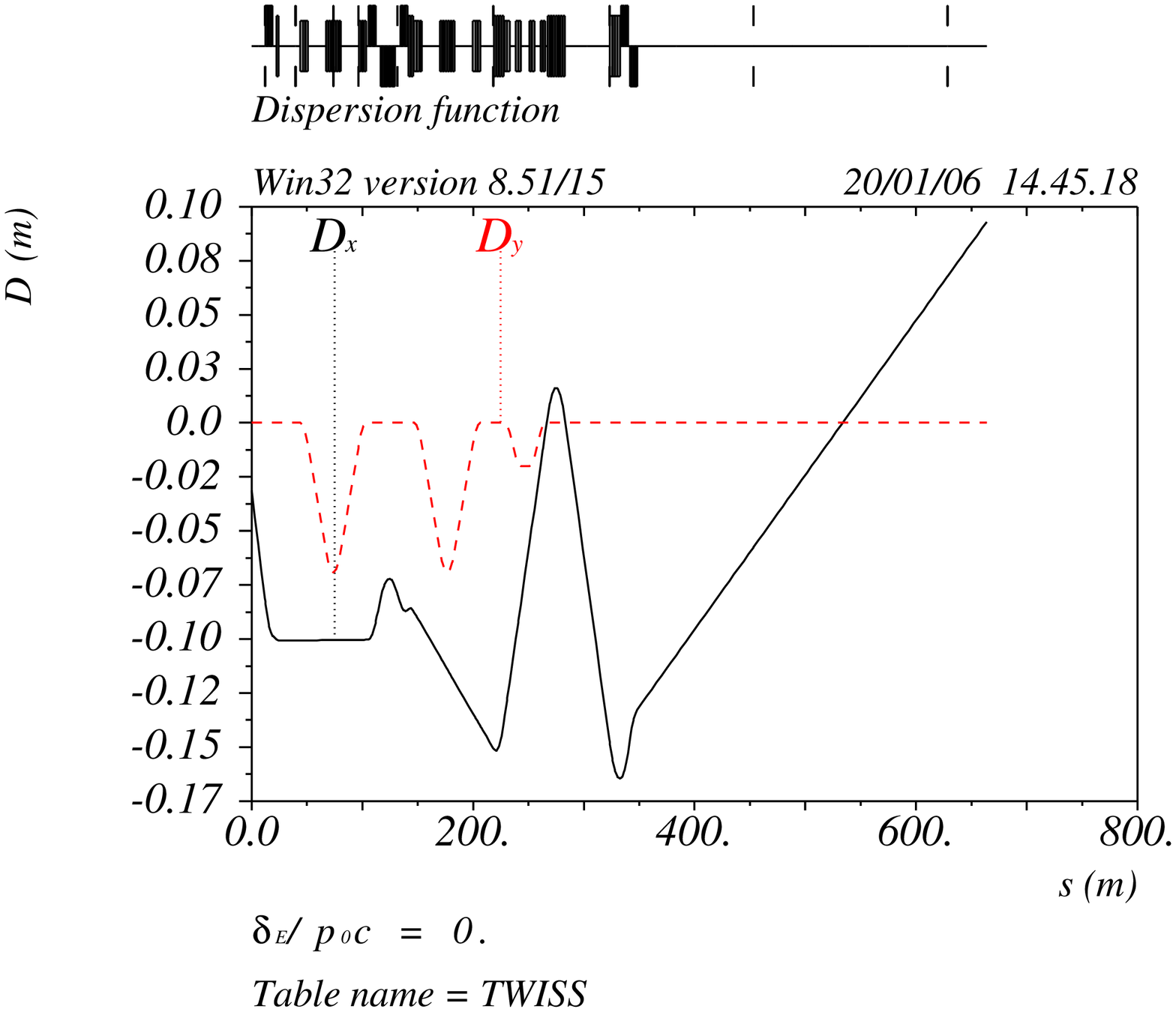}
\vspace*{-0,4cm}
\caption{\it Betatron functions along the ILC 2~mrad post-collision line 
(top) and dispersion functions downstream of the final doublet of the 
incoming beam line (bottom).}
\label{fig2mradoptics}
\vspace*{-0,3cm}
\end{figure}

\noindent
The version of the linear optics used for this study is shown in Figure~\ref{fig2mradoptics}.

\begin{figure}[h!]
\centering
\includegraphics*[width=90mm, height=90mm]{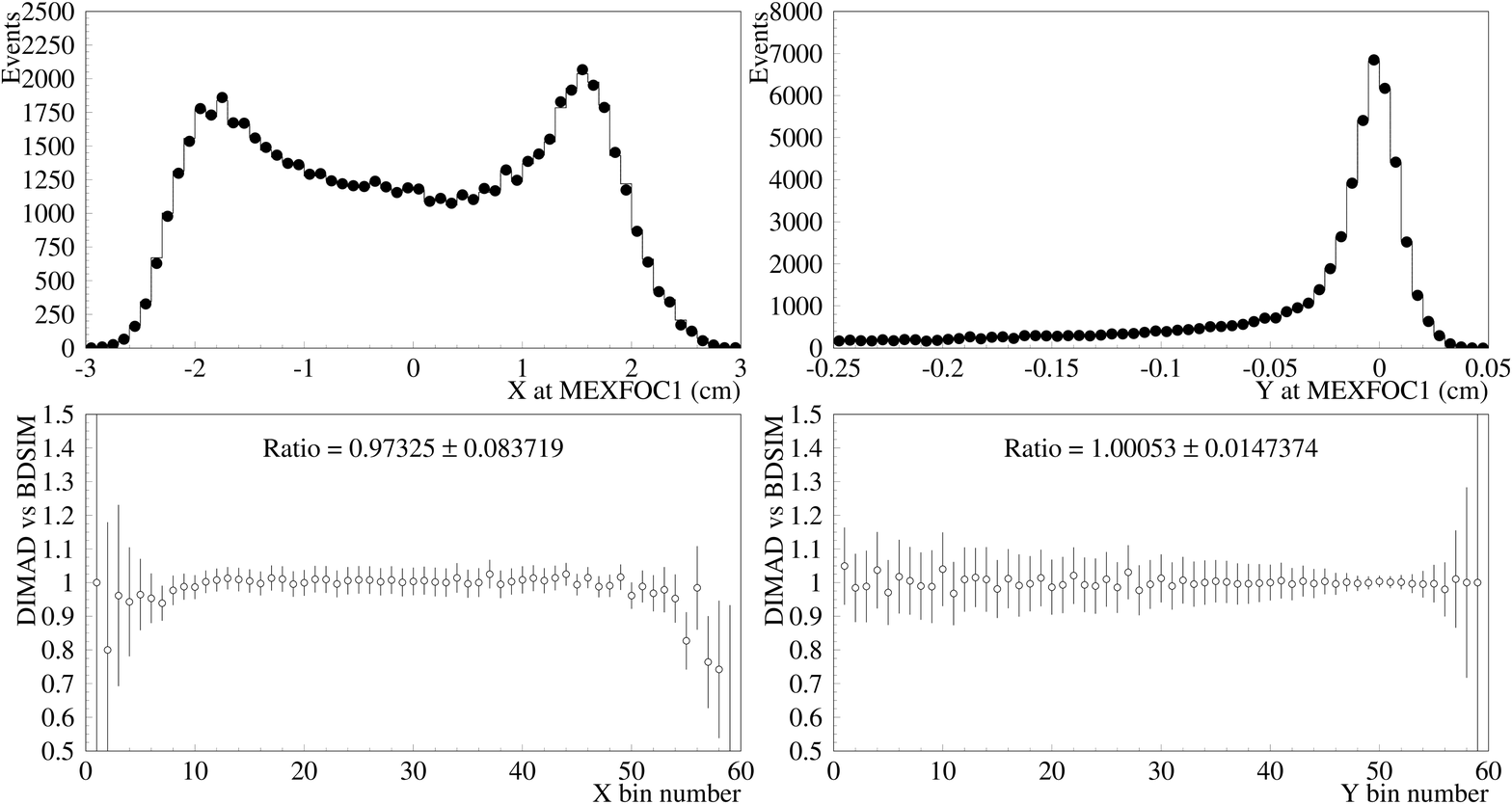}
\caption{\it Comparison of the transverse beam distributions obtained with 
DIMAD (full circles) and BDSIM (full line) at Mexfoc1. Both upper plots 
are distributed over 60 bins. The bottom plots show the ratio of the 
DIMAD and BDSIM distributions.}
\label{mexfoc1}
\end{figure}

\begin{figure}[h!]
\centering
\includegraphics*[width=90mm, height=90mm]{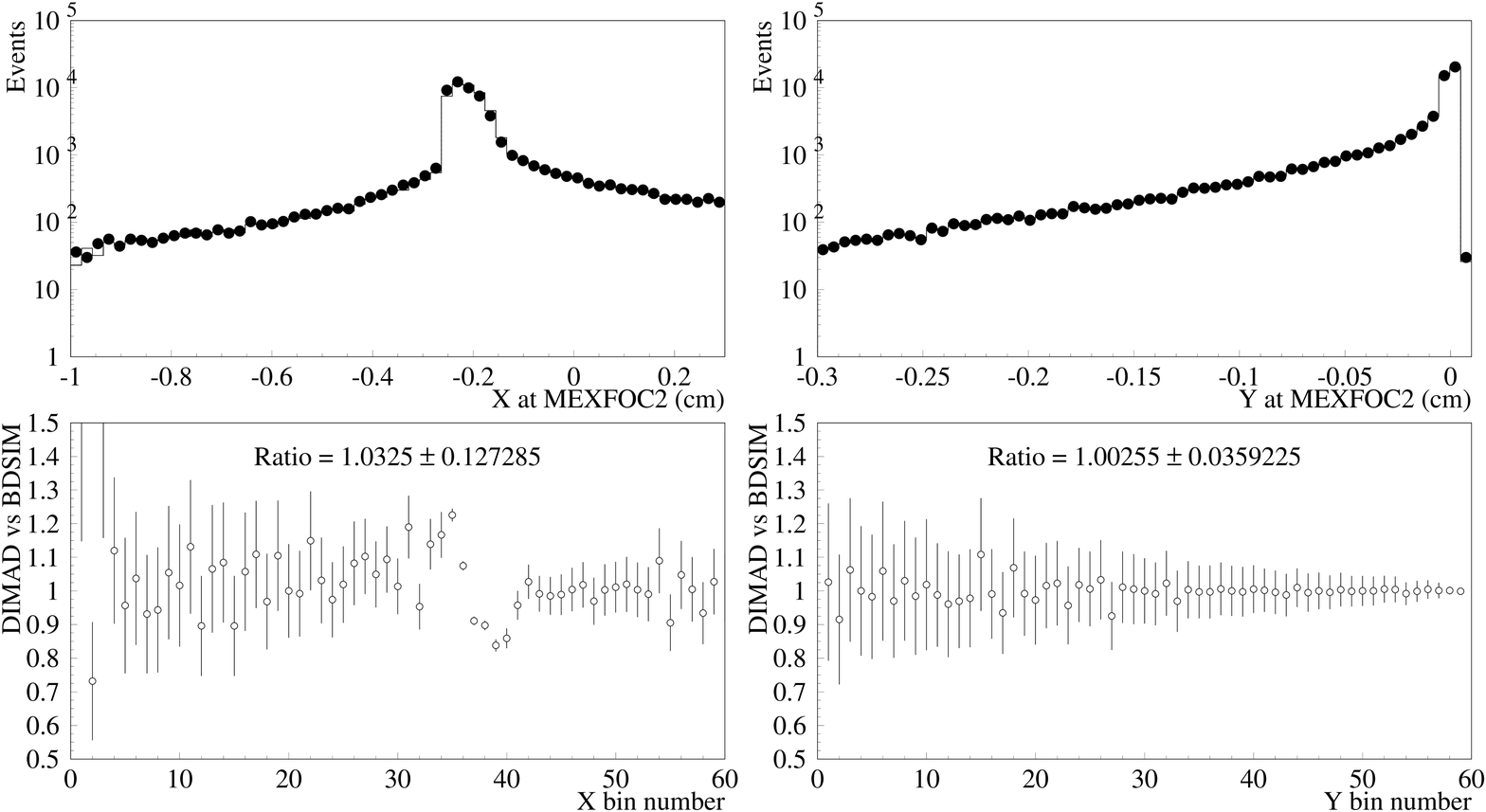}
\caption{\it Same as Figure~\ref{mexfoc1}, obtained at Mexfoc2.}
\label{mexfoc2}
\end{figure}

\begin{figure}[h!]
\centering
\hspace*{-0,7cm}\includegraphics*[width=90mm, height=90mm]{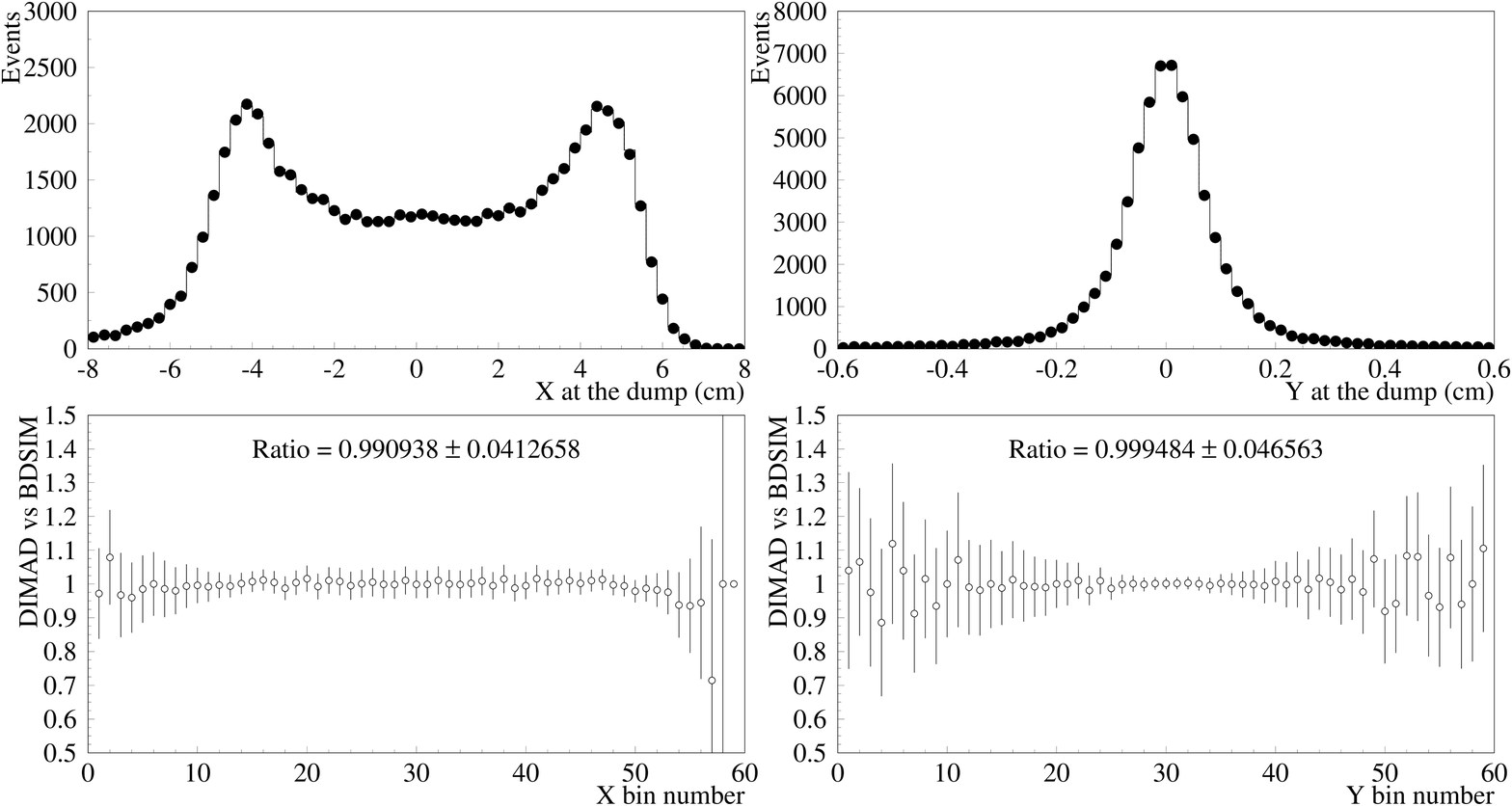}
\vspace*{0,3cm}
\caption{\it Same as Figure~\ref{mexfoc1}, obtained at the dump.}
\label{dump2}
\end{figure}

Following the BDSIM/DIMAD comparison in 20~mrad scheme, we now simply 
describe the results obtained for the 2~mrad case, using exactly the 
same beam. Figures~\ref{mexfoc1} to~\ref{dump2} show the results of 
the comparison at three locations along the 2~mrad extraction line. 
Figures~\ref{mexfoc1} and~\ref{mexfoc2} show the comparison at Mexfoc1 
(located just after the energy clean-up chicane) and at Mexfoc2 (the secondary 
focus of the polarimetry chicane), respectively. As for Figure~\ref{dump2}, it 
shows the comparison at the beam dump.

In these plots, we have projected the transverse beam distributions 
obtained from the particle tracking into bins, and we have then formed 
the ratio of the DIMAD prediction to the BDSIM prediction. As in the 
previous section, the open circles show the ratio, with the error bars 
accounting for the limited number of events in a given bin. 
All diagrams show a good agreement between DIMAD and BDSIM for 
the ILC 2~mrad extraction line, except at the secondary focus of 
the polarimetry chicane (Mexfoc2), where slight discrepancies 
are visible. These may be due to different treatments of the high-order 
effects in the optical transport through non-linear elements.

\section{CONCLUSION}

In this paper, we performed a detailed benchmarking study of two particle 
tracking codes, DIMAD and BDSIM. For this purpose, we have considered 
the ILC extraction lines with a crossing angle of 2~mrad or 20~mrad 
and, in each of these two configurations, we have performed tracking 
studies of disrupted post-collision electron beams. Here, only 
the nominal luminosity case of the 500~GeV machine was studied. We find 
that both programs give an equivalent description of the beam transport 
in all parts of the post-collision lines, except at the secondary focus 
for the 2~mrad design, where a small difference is visible.
\vspace*{0,3cm}

\end{document}